\documentclass[prb,aps,twocolumn,notitlepage]{revtex4-1}
\usepackage{amsmath}
\usepackage{amsfonts}
\usepackage{graphicx}
\usepackage[dvipsnames]{xcolor}

\begin{document}

\title {Majorana Bound States in Double Nanowires 
with Reduced Zeeman Thresholds \\ due to Supercurrents}
\author{Olesia Dmytruk, Manisha Thakurathi, Daniel Loss, and Jelena Klinovaja}
\affiliation{Department of Physics, University of Basel, Klingelbergstrasse 82, CH-4056 Basel, Switzerland}
\date{\today}

\begin{abstract}
We study the topological phase diagram of a setup composed of two nanowires with strong Rashba spin-orbit interaction subjected to an external magnetic field and brought into the proximity to a bulk $s$-wave superconductor in the presence of a supercurrent flowing through it. The supercurrent reduces the critical values of the Zeeman energy and crossed Andreev superconducting pairing required to reach the topological phase characterized by the presence of one Majorana bound state localized at each system end. We demonstrate that,  even in the regime of the crossed Andreev 
pairing being smaller than the direct  proximity pairing, a relatively weak magnetic field drives the system into the topological phase due to the presence of the supercurrent.
\end{abstract}
\maketitle

\section{Introduction}

The idea of realizing topological phases of matter such as Majorana bound states (MBSs) in condensed matter setups has been attracting substantial attention in recent years.  Zero-energy MBSs occur in topological superconductors and obey non-Abelian statistics, which makes them potential building blocks for a topological quantum computer~[\onlinecite{kitaev2001,alicea2012,beenakker2013}]. This can be achieved, for example, by coupling them to quantum dots to provide missing quantum gates~[\onlinecite{hoffman2016universal,plugge2016roadmap,karzig2017scalable}]. 
One of the most well-studied platforms for realizing MBSs is based on the  proposal of  semiconducting nanowires (NWs) with strong spin-orbit interaction (SOI) proximitized with an $s$-wave  superconductor and subjected to an external magnetic field~[\onlinecite{alicea2010majorana,lutchyn2010majorana,oreg2010helical,potter2011majorana,
sticlet2012spin,halperin2012adiabatic,san2012ac,chevallier2012mutation,klinovaja2012transition,
maier2014majorana,thakurathi2015majorana,dmytruk2015cavity}]. 
Zero-bias conductance peaks consistent with the predicted signatures of the MBSs were observed in such setups experimentally [\onlinecite{mourik2012signatures,das2012zero,
deng2012anomalous,churchill2013superconductor,finck2013anomalous}].  However, such zero-bias conductance peaks can also arise due to, for example, the presence of Andreev bound states~[\onlinecite{kells2012near,ptok2017controlling,setiawan2017electron,moore2018two,
fleckenstein2018decaying,aseev2018lifetime,vuik2018,avila2018,reeg2018zero}], which motivates the search for alternative systems such as, for example, atomic chains [\onlinecite{nadj2013proposal,klinovaja2013topological,braunecker2013interplay,
vazifeh2013self,pientka2013topological,poyhonen2014majorana,nadj2014observation,ruby2015end,
pawlak2016probing}].

\begin{figure}[b!] 
\includegraphics[width=0.8\linewidth]{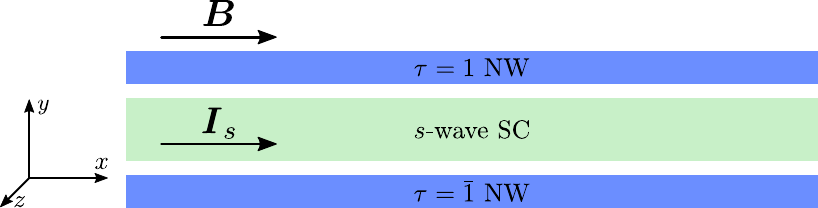}
\caption{Schematics of the double-NW setup consisting of two semiconducting Rashba NWs (blue strips), aligned along ${x}$-axis, placed in proximity to a bulk $s$-wave superconductor (green strip). The two NWs are labeled by the index $\tau= 1 \,(\bar{1})$ corresponding to the upper (lower) NW. An external magnetic field $\boldsymbol B$ is applied along the ${x}$-axis and the SOI vectors $\boldsymbol {\alpha}_{\tau}$ point along the ${z}$-axis in both NWs. The supercurrent $\boldsymbol{I}_s$ along ${x}$-axis is generated in the bulk $s$-wave superconductor, which causes a spatial gradient in the phases of the proximity-induced direct and crossed Andreev pairing amplitudes.
} 
\label{fig:scheme}
\end{figure}

One of the main challenges in realizing MBSs in 1D NWs is the need to apply relatively strong magnetic fields in order to enter the topological phase. First, magnetic fields have detrimental effects on the bulk superconductor needed to proximitize the NWs.  Second, if the coupling between a NW and a bulk superconductor is strong, an initially large $g$-factor in the NW is strongly suppressed~[\onlinecite{potter2011engineering,tkachov2013suppression,zyuzin2013correlations,cole2015effects,
van2016conductance,reeg2017transport,reeg2018metallization,reeg2018proximity,
mikkelsen2018hybridization,woods2018effective,antipov2018effects}]. 
Moreover, the orbital effects due to the applied magnetic field~[\onlinecite{lim2013emergence,nijholt2016orbital,dmytruk2018suppression}], usually neglected,  also start to play a role, making it more difficult to achieve the topological phase. To mitigate these conflicting requirements on the magnetic field,  alternative setups based on two semiconducting NWs [\onlinecite{gaidamauskas2014majorana,klinovaja2014time,thakurathi2018majorana,klinovaja2014kramers,
ebisu2016theory}], as shown in Fig.~\ref{fig:scheme}, have been proposed to realize Kramers pairs of MBSs~[\onlinecite{nakosai2013majorana,zhang2013time,keselman2013inducing,haim2014time,
dumitrescu2014magnetic,orth2015non,klinovaja2015fractional,izumida2017topology,
fleckenstein2019z}] in the absence of magnetic fields. In such a double-NW setup, there are two types of proximity effects induced by the bulk superconductor: the direct pairing with the amplitude $\Delta$ and the crossed Andreev pairing with the amplitude $\Delta_c$~[\onlinecite{deutscher2000coupling,recher2001andreev,recher2002superconductor,bena2002quantum,
hofstetter2009cooper,das2012high,deacon2015cooper}]. Only if $\Delta_c > \Delta$, the system hosts a Kramers pair of MBSs appearing at each system end. However, to achieve such a regime, strong electron-electron interactions are required [\onlinecite{thakurathi2018majorana}]. We note that crossed Andreev pairing also lies the foundations for Cooper pair splitters [\onlinecite{recher2001andreev}], which recently implemented experimentally in double-NW setups~[\onlinecite{baba2018cooper}], opening a path for studying topological properties of such systems.
Moreover, the presence of crossed Andreev pairing, even if it is still weaker than the direct pairing, can help to reduce the critical magnetic field needed to enter the topological phase [\onlinecite{schrade2017lowfield}]. Similarly,
a supercurrent flowing through the bulk superconductor was proposed as a means
allowing smaller magnetic field values
in the single NW setup ~[\onlinecite{romito2012manipulating,sticlet2013josephson,mahyaeh2018zero,lobos}]
or in an array of magnetic adatoms deposited on the surface of a bulk $s$-wave superconductor~[\onlinecite{heimes2014majorana},\onlinecite{rontynen2014tuning}].

In this work, we consider a double-NW setup consisting of two parallel NWs with Rashba SOI in proximity to  a bulk $s$-wave superconductor, see Fig. \ref{fig:scheme}.  The magnetic field is applied parallel to the NWs and perpendicular to the SOI vectors. A supercurrent is induced in the bulk superconductor that gives rise to a spatial gradient of the phase of the direct as well as of the  crossed Andreev pairing gap. This phase gradient breaks time-reversal symmetry in the system even at zero magnetic field. Thus, our setup cannot host  Kramers pairs of MBSs anymore. In general, we find that the supercurrent flowing in the bulk superconductor lowers the critical values of the Zeeman energy and crossed Andreev pairing for realizing the topological phase with one MBS at each end of the double-NW setup. Moreover, in the presence of the supercurrent the one-MBS phase can be entered at very low values of the magnetic field even in the regimes where the crossed Andreev pairing amplitude is smaller than the direct  one.

The paper is organized as follows. In Sec.~\ref{sec:model}, we present a model of our setup consisting of two parallel NWs with Rashba SOI in proximity to a bulk superconductor. We study the effect of the supercurrent on the topological phase diagram  in Sec.~\ref{sec:effectphasegradient}. We take into account the finite interwire tunneling and demonstrate that it can be compensated by the appropriate tuning of the chemical potential in Sec.~\ref{sec:effectinterwire}. The effect of a finite chemical potential on the topological phase diagram is discussed in Sec.~\ref{sec:phasechemicalpotential}. We summarize our results in Sec.~\ref{sec:conclusion}.

\section{Model for Double NW} \label{sec:model}
We consider a double-NW system consisting of two parallel semiconducting Rashba NWs labeled by the index $\tau=1$ (upper NW) and $\tau=\bar 1$ (lower NW), 
  see Fig.~\ref{fig:scheme}. The NWs, aligned along the $x$-axis, are proximity coupled to the same bulk $s$-wave superconductor. Both NWs are subjected to an external magnetic field $\boldsymbol{B}$ that is applied along the axis of the NWs. The Rashba SOI vector ${\boldsymbol {\alpha}}_{\tau}$ points along the ${z}$-axis. In an effective one-dimensional continuum model, this double-NW setup is described by the Hamiltonian~[\onlinecite{klinovaja2014time},\onlinecite{schrade2017lowfield}]
\begin{align}
 H = H_{kin} + H_{so} + H_Z  + H_{d} + H_{c}+H_\Gamma,
 \end{align}
where  the individual terms correspond to the kinetic energy, the SOI, the Zeeman energy, the direct and crossed Andreev pairing, and the interwire tunneling, respectively. 
Below, we describe each of these terms in detail.  The kinetic energy is given by
\begin{align}
H_{kin} = \sum_{\tau,\sigma}\int dx\ \psi^\dag_{\tau\sigma} (x)\left[\dfrac{(-i\,\hbar\,\partial_x)^2}{2m} - \mu_\tau\right]\psi_{\tau\sigma}(x),
\label{eq:kinetic}
\end{align}
where $\psi_{\tau\sigma}(x)$ is the annihilation operator acting on an electron of band mass $m$ and spin projection $\sigma=\pm 1$ along the $z$-axis located at position $x$ in the $\tau$--NW. Here, $\mu_\tau$ is the chemical potential of the $\tau$--NW.
 The SOI  term is given by
\begin{align}
H_{so} = i\sum_{\tau,\sigma,\sigma'}\alpha_\tau\int dx\ \psi^\dag_{\tau\sigma} (x)\left(\sigma_z\right)_{\sigma\sigma'}\partial_x\psi_{\tau\sigma'}(x),
\label{eq:soi}
\end{align}
with $\alpha_\tau$ being the SOI strength in the $\tau$--NW and $\sigma_{x,y,z}$ being the Pauli matrices acting on spin space. The energy spectrum for different spin components is split by the SOI, giving rise to two spin-polarized bands $E_{\tau \sigma}= \hbar^2 \,(k-\sigma\, k_{so,\tau})^2/2m$, with the SOI wavevector $k_{so,\tau}=m \alpha_\tau/\hbar^2$. The chemical potential $\mu_\tau$ is tuned close to the crossing point of the two spin-polarized bands at zero momentum [\onlinecite{noSOI1}], i.e. to the spin-orbit energy $E_{so,\tau}=\hbar^2 \, k_{so,\tau}^2/2m$. 
The third term in the Hamiltonian is the Zeeman term,
\begin{align}
H_Z = \sum_{\tau,\sigma,\sigma'} \Delta_{Z\tau} \int dx\ \psi^\dag_{\tau\sigma} (x)\left(\sigma_x\right)_{\sigma\sigma'}\psi_{\tau\sigma'}(x),
\label{eq:zeeman}
\end{align}
where $\Delta_{Z\tau}=g_\tau\mu_B B/2$ is the Zeeman energy, with $g_\tau$ the $g$-factor of the  $\tau$--NW and $\mu_B$  the Bohr magneton. 

In the double-NW, as discussed previously~[\onlinecite{klinovaja2014time},\onlinecite{thakurathi2018majorana},\onlinecite{schrade2017lowfield},\onlinecite{ reeg2017diii}], we have two types of proximity-induced superconductivity giving rise to direct [$\Delta_\tau(x)$] and crossed Andreev [$\Delta_c (x)$] pairing amplitudes. The direct (intrawire) pairing term $H_{d}$ finds its origin in tunneling of both electrons of a Cooper pair into one of the two NWs, and it is given by
\begin{align}
H_{d} = \dfrac{1}{2}\sum_{\tau,\sigma,\sigma'}\int dx\ \left[\Delta_\tau(x) \psi_{\tau\sigma} (x)\left(i\sigma_y\right)_{\sigma\sigma'}\psi_{\tau\sigma'}(x) + \text{H.c.}\right].
\label{eq:directpairing}
\end{align}
If the two electrons of a Cooper pair get split and tunnel into two different NWs, the crossed Andreev pairing (interwire) term is generated and given by
\begin{align}
H_{c} = \dfrac{1}{2}\sum_{\tau,\sigma,\sigma'}\int dx\ \left[\Delta_c(x) \psi_{\tau\sigma} (x)\left(i\sigma_y\right)_{\sigma\sigma'}\psi_{\bar\tau\sigma'}(x) + \text{H.c.}\right].
\label{eq:crossedandreevpairing}
\end{align}
In the presence of a supercurrent $\boldsymbol {I}_s$ flowing along the $x$-axis in the bulk $s$-wave superconductor (see Fig.~\ref{fig:scheme}) both the direct and crossed Andreev pairing amplitudes acquire a position-dependent phase $\varphi(x)$~[\onlinecite{romito2012manipulating}],
\begin{align}
&\Delta_\tau (x) = \Delta_\tau e^{-i\varphi(x)},\\
&\Delta_c (x) = \Delta_c e^{-i\varphi(x)},
\end{align}
where $\varphi(x)$ is the same for all pairing amplitudes. 
In the following, we work with a uniform supercurrent  such that the phase $\varphi(x)$ changes linearly as a function of position $x$ and has a form $\varphi(x) = x/\xi$, where $\xi$ is a characteristic lengthscale. 
We assume that the supercurrent ($I_s \propto 1/\xi$) stays always smaller than the critical current ($I_c \propto  1/\xi_{sc}$) in the bulk $s$-wave superconductor characterized by  the coherence length $\xi_{sc} \ll \xi$ ~[\onlinecite{romito2012manipulating}].

Finally, we also consider the interwire tunneling, which has the following form
\begin{align}
H_{\Gamma} = -\Gamma\sum_{\tau,\sigma}\int dx\ \psi^\dag_{\tau\sigma} (x)\psi_{\bar\tau\sigma}(x),
\label{eq:interwiretunneling}
\end{align}
where $\Gamma$ is the tunneling strength. In what follows, we assume $\Delta_{Z\tau}$, $\Delta_\tau$, $\Delta_c$, and $\Gamma$ to be real and non-negative without loss of generality. For simplicity, we also assume that $\Delta_1 = \Delta_{\bar 1} \equiv \Delta$ and $\Delta_{Z,1} = \Delta_{Z,\bar 1} \equiv \Delta_Z$. 
In terms of the  Pauli matrices $\tau_{x,y,z}$, $\eta_{x,y,z}$, and $ \sigma_{x,y,z}$, which act in the NW, particle-hole, and spin spaces, respectively, we write for the Hamiltonian, $H=\int dx\, \Psi^\dagger \mathcal{H} \Psi/2$,  where the Hamiltonian density $\mathcal{H}$  has the  form
\onecolumngrid
\begin{align}
\mathcal{H} &= -\dfrac{\hbar^2\partial_x^2}{2m}\eta_z -\mu_1 \dfrac{1+\tau_z}{2}\eta_z -\mu_{\bar 1} \dfrac{1-\tau_z}{2}\eta_z\notag
+i\alpha_1\dfrac{1+\tau_z}{2}\sigma_z\partial_x
+i\alpha_{\bar 1}\dfrac{1-\tau_z}{2}\sigma_z\partial_x -\Gamma \tau_x \eta_z\notag\\
&\hspace{95pt}+ \Delta_Z \eta_z\sigma_x  -i \Delta\Big[e^{i\varphi(x)}\eta_+\sigma_y - e^{-i\varphi(x)}\eta_-\sigma_y\Big] -i\Delta_c\Big[e^{i\varphi(x)}\tau_x\eta_+\sigma_y - e^{-i\varphi(x)}\tau_x\eta_-\sigma_y\Big].
\label{eq:doublewiregeneral}
\end{align}
\twocolumngrid
In what follows, we use the basis $\Psi$=
($\psi_{1\uparrow}$, $\psi_{1\downarrow}$, $\psi_{1\uparrow}^\dagger$, 
$\psi_{1\downarrow}^\dagger$, $\psi_{2\uparrow}$, $\psi_{2\downarrow}$, $\psi_{2\uparrow}^\dagger$,$\psi_{2\downarrow}^\dagger )^T$.  Here,  we have introduced the notation $\eta_{\pm}=(\eta_x \pm i\eta_y)/2$. To simplify analytical calculations,  it is convenient to get rid of the spatially-periodic phase $\varphi(x)$ in the  pairing terms. This can be achieved  by 
a unitary transformation $U = e^{i \varphi(x) \eta_z / 2}$~[\onlinecite{romito2012manipulating}],
leading to the new Hamiltonian density $\mathcal{H}' = U^\dag \mathcal{H} U$ that restores the translation invariance.
As a result,  $\mathcal{H}'$ can be easily rewritten in momentum space representation as 
\onecolumngrid
\begin{align}
&\mathcal{H}'(k) = \dfrac{\hbar^2}{2m}\left(k + \dfrac{\eta_z}{2\xi}\right)^2\eta_z -\mu_1 \dfrac{1+\tau_z}{2}\eta_z -\mu_{\bar 1} \dfrac{1-\tau_z}{2}\eta_z
- \alpha_1\left(k + \dfrac{\eta_z}{2\xi}\right)\dfrac{1+\tau_z}{2}\sigma_z \notag\\
&\hspace{95pt}- \alpha_{\bar 1}\left(k + \dfrac{\eta_z}{2\xi}\right)\dfrac{1-\tau_z}{2}\sigma_z-\Gamma \tau_x \eta_z + \Delta_Z \eta_z\sigma_x  + \Delta \eta_y\sigma_y + \Delta_c\tau_x \eta_y\sigma_y.
\label{eq:generalmomentum}
\end{align}
\twocolumngrid

In the absence of a supercurrent and at zero magnetic field, the Hamiltonian $\mathcal{H}'$ is time-reversal symmetric, $T^{\dag} \mathcal{H}' (k) T = \mathcal{H}'(-k)$, where $T= -i \sigma_y K$ and $K$ is the complex conjugation operator. The finite phase gradient $\partial_x \varphi(x)$ breaks time-reversal symmetry in such a way that, in contrast to Ref.~[\onlinecite{schrade2017lowfield},\onlinecite{reeg2017diii}], 
the effective time-reversal symmetry operator cannot be defined.
The Hamiltonian $\mathcal{H}'$ still obeys particle-hole symmetry, $P^\dag\mathcal{H}'(k)P = -\mathcal{H}'(-k)$, where  $P = \eta_x$ is the particle-hole operator. As a result, $\mathcal{H}'$ can be classified to belong to  the topological symmetry class D characterized by a $\mathbb{Z}_2$ topological invariant~[\onlinecite{ryu2010topological}].

\section{Topological phase diagram} \label{sec:phasediagram}

In this section, we study the topological properties of the double NW  in the presence of a supercurrent and crossed Andreev pairing. 
We demonstrate that, depending on the values of the parameters, the system can be (1) in a gapped trivial phase hosting no bound states (BSs) or hosting one fermion bound state (FBS), (2) in a gapless phase, and (3) in a gapped topological phases hosting one or two MBSs at each end of the setup. The number of MBSs, being a topological invariant, cannot change without closing of the bulk gap of the double-NW system, therefore, the topological phase transition is associated with closing and reopening of the gaps in the energy spectrum. Thus, potential boundaries between different gapped phases can be identified by the bulk gap closing at zero momentum.
However, one needs to check explicitly if the system is still gapped for a given set of parameters.  The transition to the gapless phase occurs due to the bulk gap closing at finite momentum and can be found, in general, only numerically. Thus, in addition to our analytical calculations, we study the phases also numerically by diagonalizing the corresponding `tight-binding' model [i.e. the discretized version of $\mathcal{H}'$ in real space] with  hopping amplitude $t = \hbar^2/2 m a^2 = 25$~meV, where $m = 0.015 m_e$ is the effective mass, $a = 10$ nm the effective lattice constant, and $N=1000$ number of sites.  

Our main focus is the topological phase with one MBS, which we aim to achieve at the lowest possible strengths of the crossed Andreev pairing amplitude $\Delta_c$ and the Zeeman energy $\Delta_Z$.
We note that the topological phase with two topologically protected MBSs occurs only in two special cases. First, in the absence of the supercurrent and magnetic field, the system is time-reversal invariant and hosts Kramers pair of MBSs, if the crossed Andreev pairing dominates over the direct one [\onlinecite{klinovaja2014time},\onlinecite{schrade2017lowfield}]. Second, in the absence of any coupling between the two NWs, i.e. in the absence of crossed Andreev pairing and interwire tunneling, the two NWs can be treated independently. If both of the NWs are in a topological phase hosting one MBS each, together they host two MBSs [\onlinecite{schrade2017lowfield}], which are protected  from any hybridization due to their spatial separation. However, if in such a system one adds a weak coupling between the two NWs  as caused by $\Delta_c$ or $\Gamma$, these two MBSs at the same end of the system hybridize into a single FBS. 
We also note that the gapless phase, characterized by zero-energy bulk modes at finite momentum,  can emerge only in the presence of a phase gradient~[\onlinecite{romito2012manipulating}] or of strong magnetic fields suppressing the superconducting pairings. Obviously, there are no BSs in such a gapless regime and it can, thus, be considered as a trivial phase.

\subsection{Effect of the phase gradient: topological phase at smaller values of Zeeman energy and of crossed Andreev pairing} \label{sec:effectphasegradient}

The main goal of this work is to demonstrate that the topological phase characterized by the presence of one MBS at each end of the system can be achieved at much lower magnetic fields if one induces supercurrents in the $s$-wave bulk superconductor. To shed light on this effect, we first consider the simplest analytical model given by Eq.~\eqref{eq:generalmomentum}, where
the chemical potential is tuned to the crossing of the spin-polarized bands at $k=0$ and, thus, we set $\mu_\tau = 0$ in Eq.~\eqref{eq:generalmomentum}. 
The case with $\mu_\tau \neq 0$ will be discussed further below in following sections.
Also, for the moment, we assume that there is no interwire tunneling, $\Gamma = 0$, as it was demonstrated before that any finite $\Gamma$ can be compensated by tuning the chemical potential to a sweet-spot~[\onlinecite{schrade2017lowfield}]. In the next section, we  demonstrate that such a compensation can be achieved also for the present setup. We also focus first on the generic case $\alpha_1\neq\alpha_{ \bar 1}$ and comment on the special case $\alpha_1=\alpha_{\bar 1}$ in the next subsection.

\begin{figure}[t!] 
\includegraphics[width=0.8\linewidth]{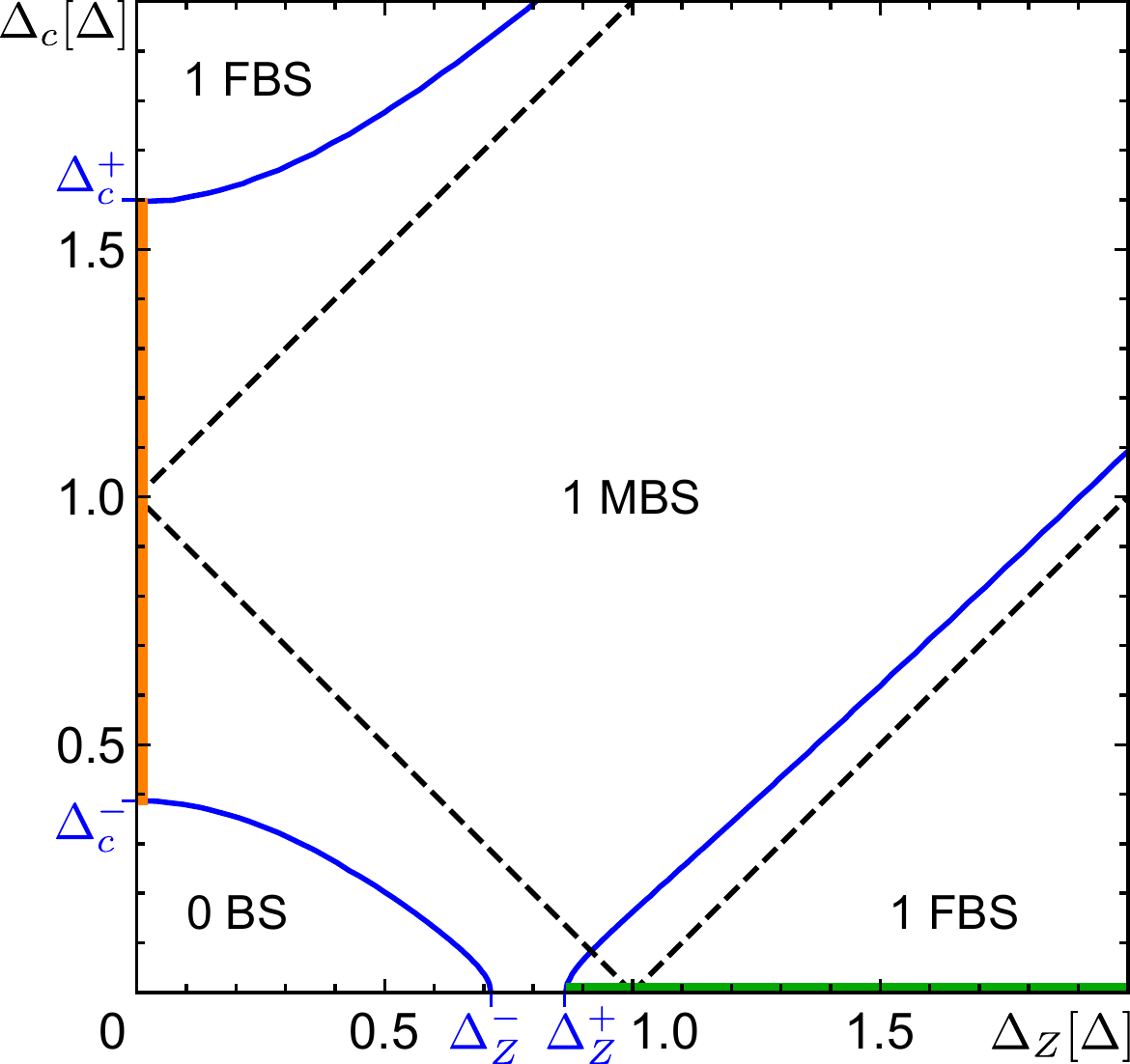}
\caption{
Phase diagram of the double-NW system as function of Zeeman energy $\Delta_Z/\Delta$  and  crossed Andreev pairing amplitude $\Delta_c/\Delta$.
The phase boundaries between different gapped phases in the absence of supercurrent ($\delta = 0$, black dashed line) and in the presence of supercurrent ($\delta/\Delta= 0.04$, blue solid line) are found from Eq.~\eqref{eq:phasediagram1}.  The number of BSs is indicated directly in the plot and refers to a given end of the double NW.  For finite supercurrent, the topological phase characterized by the presence of one MBS extends to weaker magnetic fields and weaker crossed Andreev pairings.  
For $\delta = 0$, the one-FBS phase becomes a two-MBSs phase, which is also the case for $\Delta_c=0$  and $\Delta_Z > \Delta_Z^+$ (green solid line) and finite $\delta$. 
In the absence of supercurrent, $\delta = 0$, and in the absence of magnetic field (crossed Andreev pairing), the bulk gap at $k=0$ closes at $\Delta_c=\Delta$ ($\Delta_Z=\Delta$). 
For finite $\delta$ and in the absence of magnetic field (crossed Andreev pairing), the bulk gap at $k=0$ closes twice at $\Delta_c=\Delta_c^\pm$ ($\Delta_Z=\Delta_Z^\pm$).
In the presence of supercurrent, the system is in the gapless phase at $\Delta_Z = 0$ and  $\Delta_c^- \leq  \Delta_c \leq \Delta_c^+$ (orange solid line). 
Other parameters are fixed as $\mu_1 = \mu_{\bar 1} = 0$, $\Gamma=0$, $E_{so,1}/\Delta = 6.25$, $\beta = 1.4$.} 
\label{fig:doubleNW}
\end{figure}

Analyzing Eq.~\eqref{eq:generalmomentum}, we find that the  gap in the bulk spectrum of the double NW  is closed at $k = 0$, if $\Delta_Z= \tilde \Delta_{Z,\pm}$, where $\tilde \Delta_{Z,\pm}$ is a real non-negative  solution of
\begin{align}
&\tilde \Delta_{Z,\pm}^2 = \Delta ^2+\Delta_c^2 -\left(\beta ^2+1\right)   E_{so,1}\,\delta/2 + \delta ^2/16 \label{eq:phasediagram1}\\
&\pm\sqrt{4 \,\Delta ^2 \Delta_c^2+\left(\beta ^2-1\right)^2 E_{so,1}^2\,\delta ^2 /4-(\beta -1)^2 \Delta_c^2 \,E_{so,1}\, \delta } \notag,
\end{align}
where $\beta = \alpha_{\bar 1}/\alpha_1 \geq 1$ and $\delta = \hbar^2/2\,m\,\xi^2$. In what follows, we focus on the part of the phase diagram obtained for $\Delta_Z\geq 0$ as a function of the crossed Andreev pairing $\Delta_c \geq 0$, see Fig. \ref{fig:doubleNW}. 

To begin, let us remind how the phase diagram looks like in the regime $\beta \gg 1$  and in the absence of supercurrents [\onlinecite{schrade2017lowfield}], $\delta=0$,
 see Fig. \ref{fig:doubleNW}. In this case, $\tilde \Delta_{Z,\pm} = |\Delta\pm\Delta_c|$. The system hosts one MBS at each system end, if $\tilde \Delta_{Z,-} < \Delta_Z < \tilde \Delta_{Z,+}$. Otherwise, the system is either in a trivial phase, hosting no BSs [if $\Delta_Z< \tilde \Delta_{Z,-}$ and $\Delta_c<\Delta$] or in a topological phase with two MBSs at each system end [if $\Delta_Z >\tilde \Delta_{Z,+}$ or if $\Delta_Z< \tilde \Delta_{Z,-}$ for $\Delta_c >\Delta$]. However, this two-MBSs phase, occurring due to the presence of an effective time-reversal symmetry, is unstable against changes in the magnetic field and SOI vector directions, as well as against disorder~[\onlinecite{schrade2017lowfield}] and can be referred to as a trivial phase. We also note that, for $\delta=0$ and $\Delta_Z = 0$, the gap in the bulk  spectrum can also close at a finite momentum at  $\Delta_c^*\equiv \Delta\sqrt{ 1 + \left(\beta^2  - 1\right)^2 E_{so,1}^2 /\Delta^2}$, see Ref.~[\onlinecite{schrade2017lowfield}]. 
Here, we have focused on the regime $\Delta_c\ll\Delta_c^*$ (or $\beta -1 \gtrsim  \Delta/2 E_{so,1}$)
 and we comment later on the regime of equal SOI strengths in the two NWs, {\it i.e.} $\beta\approx 1$. 
 
 Notably, in the absence of supercurrent, there is one point at each coordinate axis where all three phases come together. Such points are often referred to as bifurcation points in the phase diagram~[\onlinecite{klinovaja2012transition},\onlinecite{kennes2018chiral}].
 The first point is $\Delta_c=\Delta$ and $\Delta_Z=0$ and the second one  $\Delta_Z=\Delta$ and $\Delta_c=0$. These points describe a critical value of $\Delta_Z$ or $\Delta_c$ required to enter the topological phases in the absence of the other one.
Further below, we demonstrate that these bifurcation points disappear, if one applies a finite supercurrent, see Fig. \ref{fig:doubleNW}. In particular, in the absence of magnetic field (crossed Andreev pairing), the bulk gap at $k=0$ closes twice at $\Delta_c=\Delta_c^\pm$ ($\Delta_Z=\Delta_Z^\pm$).

First, we consider a set of parameters for which $\tilde \Delta_{Z,+}$ is always well-defined. Later, we show numerically that if it is not the case, which would correspond to large values of supercurrent, the system is in the gapless phase. At the same time, $\tilde \Delta_{Z,-}$ is not defined for some range of crossed Andreev pairings, $\Delta_c^-<\Delta_c < \Delta_c^+$.
As a result, Eq.~\eqref{eq:phasediagram1}  implicitly determines the boundaries between different gapped topological phases,  see Fig. \ref{fig:doubleNW}.
The system is in the trivial phase without any BSs inside the gap for $\Delta_Z < \tilde \Delta_{Z,-}$, provided that $\Delta_c < \Delta_c^-$. The system is in the trivial phase with one FBS localized at each double-NW end if $\Delta_Z > \tilde \Delta_{Z,+}$ or  if $\Delta_Z < \tilde \Delta_{Z,-}$, provided that $\Delta_c > \Delta_c^+$. Finally, the system is in the topological phase with one MBS localized at each end of the system, if $\tilde \Delta_{Z,-} < \Delta_Z < \tilde \Delta_{Z,+}$, provided that $\Delta_c < \Delta_c^-$ or $\Delta_c >\Delta_c^+$, or if $\Delta_Z < \tilde \Delta_{Z,+}$, provided that $\Delta_c^-<\Delta_c < \Delta_c^+$.

Next, we focus on the parameter regime  when $\xi \gg \xi_{sc}$. This allows us to introduce a small parameter, $\sqrt{E_{so,1}\delta}/\Delta \ll 1$. Under this approximation we find that 
\begin{align}
&\Delta^{\pm}_c\approx \Delta \pm (\beta +1) \sqrt{E_{so,1}\delta}/2,\label{eq:zeropoints0}\\
&\Delta_Z^- \approx \Delta - \beta^2  E_{so,1} \delta / 2\Delta, \label{eq:zeropoints1}\\
&\Delta_Z^+ \approx \Delta -  E_{so,1} \delta/2\Delta.
\label{eq:zeropoints2}
\end{align}
Here, we note that $\Delta^{\pm}_c=\Delta$ for $\delta=0$, as expected. At finite values of the supercurrent, $\Delta^{\pm}_c$ split away from $\Delta$ symmetrically. At the same time, $\Delta_Z^\pm<\Delta$.
This ensures that the boundary between the topological and trivial (without BS) phase  gets pushed down to smaller values of both Zeeman energy and crossed Andreev pairing. This means that MBSs can be observed experimentally at reduced magnetic fields and for crossed Andreev pairings that could be substantially weaker than $\Delta$, see Fig. \ref{fig:doubleNW}.

\begin{figure}[t!] 
\includegraphics[width=0.8\linewidth]{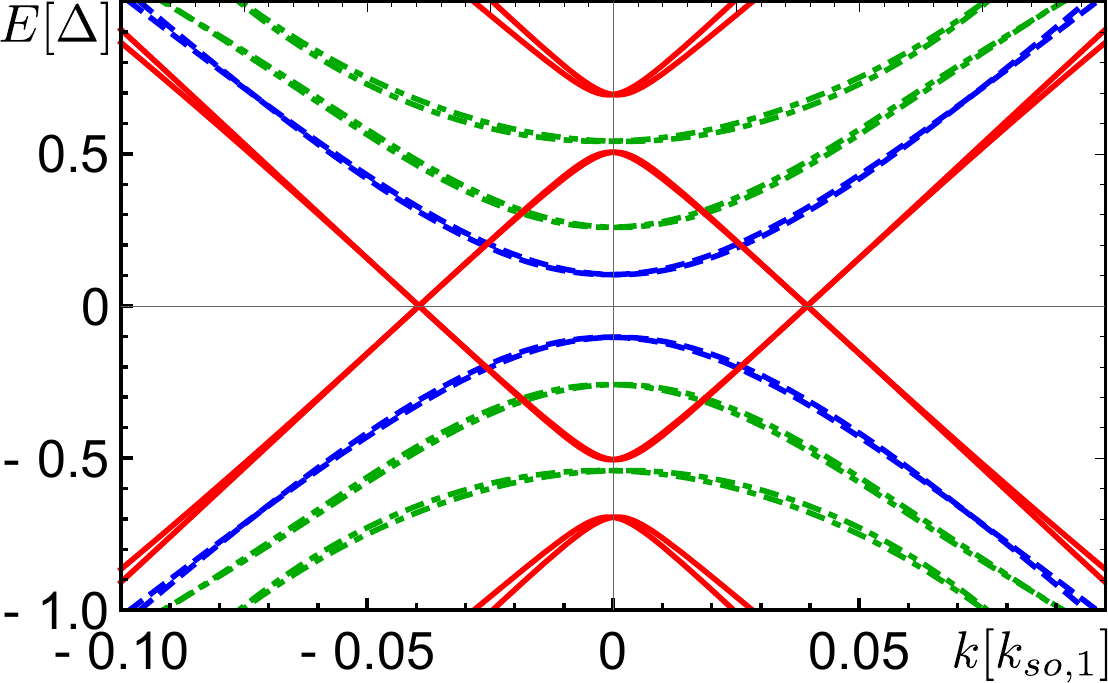}
\caption{Bulk energy spectrum $E/\Delta$ as a function of momentum $k/k_{so,1}$ for $\Delta_Z = 0$ and different values of $\Delta_c$. In the trivial phase, $\Delta_c < \Delta_c^-$ ($\Delta_c/\Delta=0.1$),  the spectrum is gapped for all momenta (green dot-dashed line). For $\Delta_c^- < \Delta_c < \Delta_c^+$ ($\Delta_c/\Delta=0.9$), the spectrum is gapped at $k = 0$ but there is no gap at some finite momentum (red solid line). Thus, the system is in the gapless phase without any BSs in the spectrum. In the one-FBS phase, $\Delta > \Delta_c^+$ ($\Delta_c/\Delta=1.7$), again, the energy spectrum is gapped for all values of $k$ (blue dashed line). Other parameters are the same as in Fig.~\ref{fig:doubleNW}.} 
\label{fig:bulk}
\end{figure}

In the above analysis, we have focused on the gap closing at $k=0$. This allowed us to find boundaries between different gapped phases. At the same time, we still need to check numerically if the system is gapped for which we need to study the spectrum for all values of momenta. 
However, for non-zero phase gradients the condition on the system being gapped can, in general, be determined only numerically [\onlinecite {romito2012manipulating}]. For example, we checked that for the system parameters used in Fig. \ref{fig:doubleNW} the system is always gapped except at the phase boundaries or along the line $\Delta_Z = 0$ and $ \Delta_{c}^-<\Delta_c < \Delta_{c}^+$, see Fig. \ref{fig:bulk}. Fixing $\Delta_Z = 0$ and increasing $\Delta_c$, the gap at $k=0$ gets smaller and smaller  until it closes at $ \Delta_{c}^-$. Increasing $\Delta_c$ further, one reopens the gap at $k=0$, however, the spectrum of the system stays gapless at some finite momentum $\pm k_g$, which first grows and later shrinks as a function of $\Delta_c$ until it comes back to $k=0$ at $\Delta_c =\Delta_c^+$. For $\Delta_c > \Delta_c^+$ we enter the gapped phase, hosting one FBS at each system end.

In the regime of substantially different SOI energies, $\beta \gg 1$, we can find analytically the values of the momentum $k_g$ for which the spectrum is gapless by linearizing the Hamiltonian $\mathcal{H}$ in Eq.~\eqref{eq:doublewiregeneral} around the Fermi points $k_F = 0, \pm 2 k_{so,\tau}$ (see Ref.~[\onlinecite{composite}]):
\begin{align}
&\left(k_g/k_{so,1}\right)^2 = \Big[2 \beta ^2 E_{so,1}\delta-2 \beta  \Delta_c^2- \left(\beta ^2+1\right) \Delta ^2\notag\\
&\hspace{20pt}+  (\beta +1) \Delta  \sqrt{(\beta -1)^2 \Delta ^2+4 \beta  \Delta_c^2}\Big]/8\beta^2 E_{so,1}^2.
\label{eq:kg}
\end{align}
In the special case  $\Delta_c = \Delta$, the gap in the bulk  spectrum closes at the maximum value of the momentum, $k_g = 1/2\xi$.

\begin{figure}[b!] 
\includegraphics[width=0.8\linewidth]{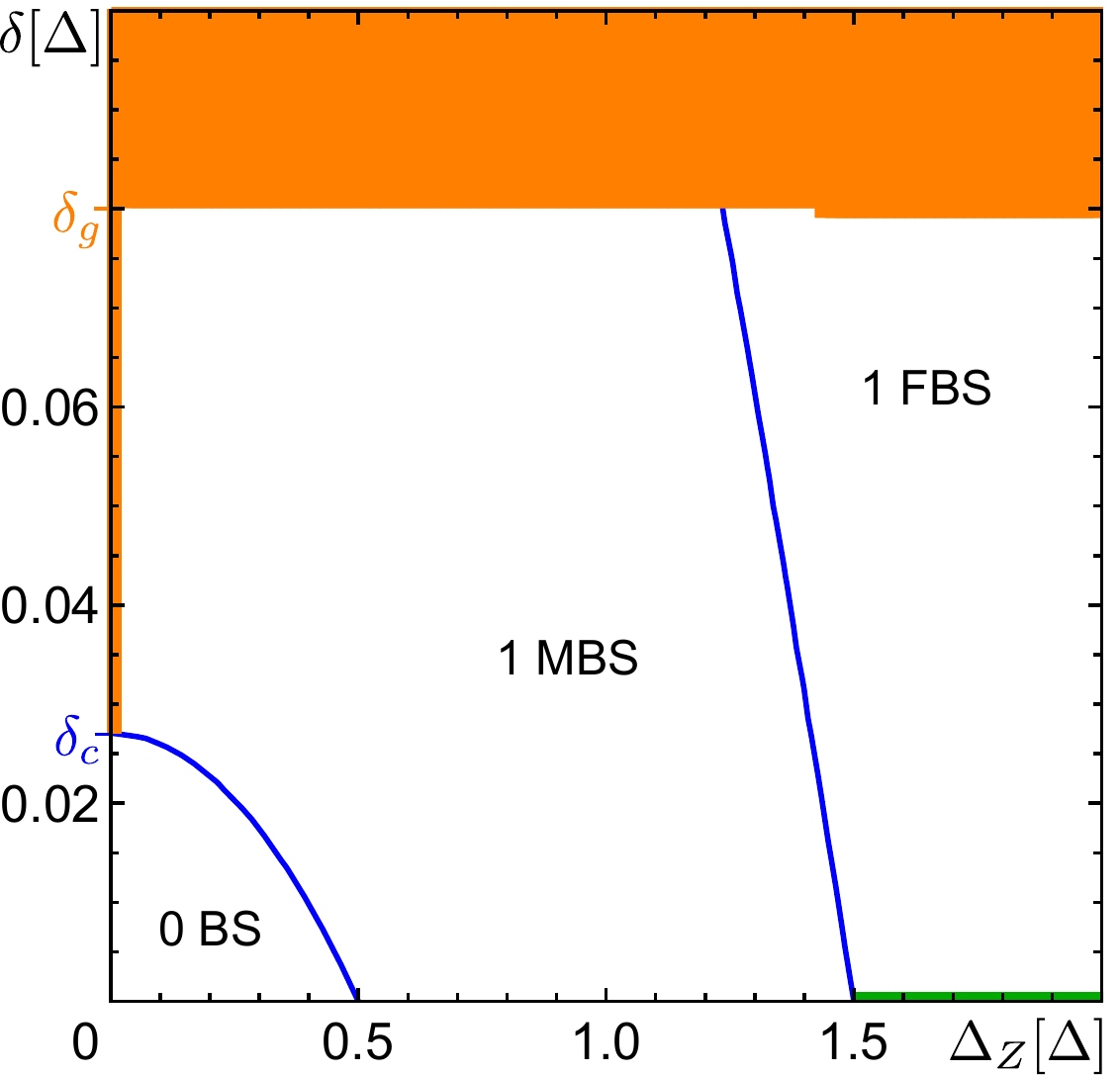}
\caption{Phase diagram of the double NW 
as function of  Zeeman energy $\Delta_Z/\Delta$ and phase gradient $\delta/\Delta$  for  fixed crossed Andreev pairing ($\Delta_c/\Delta = 0.5$).  The number of BSs in the corresponding phases is indicated directly in the figure.  If the supercurrent is too strong  $\delta>\delta_g$, the system is in the gapless phase (orange area). 
Generally, the supercurrent helps to enter the topological phase hosting one MBS at lower values of $\Delta_Z$. If $\delta_c<\delta <\delta_g$, already a relatively weak magnetic field is sufficient to achieve the topological phase, however, if $\Delta_Z = 0$, the system is in the gapless phase (orange line).
The phase boundaries between gapped phases (blue solid lines) are given by Eq.~\eqref{eq:phasediagram1}.
In the special case $\delta = 0$, a zero-energy FBS is composed of two MBSs (green line) [\onlinecite{schrade2017lowfield}].
 Other parameters are the same as in Fig.~\ref{fig:doubleNW}.} 
\label{fig:Diagramzeemanphase}
\end{figure}

As already follows from Eqs.~\eqref{eq:zeropoints0}-\eqref{eq:zeropoints2}, applying a supercurrent, we can reach the topological phase at smaller values of Zeeman energy and crossed Andreev pairing. Thus, it is beneficial to work with large values of supercurrent.
However, one should note that if the supercurrent is too strong, $\delta>\delta_g$, the system becomes gapless. 
To demonstrate this, we focus the topological phase diagram as a function of $\Delta_Z/\Delta$ and $\delta/\Delta$ for the fixed value of crossed Andreev pairing,
 see Fig.~\ref{fig:Diagramzeemanphase}. 
The phase boundaries between gapped phases in the presence of supercurrent are given by Eq.~\eqref{eq:phasediagram1} provided that $\tilde\Delta_{Z,\pm}$ are well-defined. The system is in the trivial phase with no BSs, if $\Delta_Z < \tilde\Delta_{Z,-}$, or in a one-FBS phase, if $\Delta_Z>\tilde\Delta_{Z,+}$.  Finally, the system is in the topological phase hosting one MBS at each end if $\tilde \Delta_{Z,-} < \Delta_Z < \tilde\Delta_{Z,+}$ provided that $\delta<\delta_c$ or $0 < \Delta_Z < \tilde\Delta_{Z,+}$ provided that  $\delta_c \leq \delta < \delta_g$. 
In the regime of weak supercurrent $\sqrt{E_{so,1}\delta}/\Delta \ll 1$, $\delta_c$ can be  approximated using  Eq.~\eqref{eq:zeropoints0} as $\delta_c \approx 4\left(\Delta-\Delta_c\right)^2/\left(\beta+1\right)^2E_{so,1}$.
In the absence of magnetic fields, $\Delta_Z = 0$, the spectrum is gapless for $\delta \geq \delta_c$.
However, in this regime, already a very weak magnetic field drives the system into the topological phase with one MBS. Thus,  supercurrents make it easier to generate MBSs.
However, there is also a critical value $\delta_g$ such that  for  $\delta>\delta_g$, the system becomes gapless. The precise value $\delta_g$ can be found only numerically, however, we can estimate it to be such that $\sqrt{E_{so,1}\delta_g}/\Delta =O(1)$.

To conclude, we find that in the presence of a supercurrent in the system, the critical values of the Zeeman energy and crossed Andreev pairing amplitudes for entering the topological phase characterized by the presence of one MBS at each system end could be substantially reduced. 

\subsection{Effects of interwire tunneling}\label{sec:effectinterwire}

In this subsection, we take into account the finite interwire tunneling $\Gamma$, which was neglected so far, and study its effect on the topological phase diagram.  We show that  the effect of the interwire tunneling at small values of the Zeeman energy can be compensated by tuning the chemical potential to $ \mu_\tau = \Gamma $ [\onlinecite{schrade2017lowfield}].

First, we discuss the case of different SOI strengths in both NWs, $\beta \gtrsim 1 + \Delta/2 E_{so,1}$. The phase diagram for the case $\mu_\tau=\Gamma=0$ was described in detail in the previous subsection, see Fig. \ref{fig:doubleNW}. 
Here, we assume the chemical potentials in both NWs to be tuned to the interwire tunneling $\Gamma$ such that $\mu_1 = \mu_{\bar 1}=\Gamma$. 
The gap at $k = 0$ in the bulk spectrum is closed, if $\Delta_Z=\tilde{\tilde{\Delta}}_{Z,\pm}$ , where $\tilde{\tilde{\Delta}}_{Z,\pm}$ is a real non-negative solution of
\begin{align}
&\tilde{\tilde{\Delta}}_{Z,\pm}^2 = \Delta ^2+\Delta_c^2 -\delta  \left[\Gamma +(\beta ^2+1) E_{so,1}\right]/2  +\delta ^2/16\notag\\
&\hspace{0pt}+ 2 \Gamma ^2 \pm\Big\{4 \Gamma ^4 - 2 \Gamma ^3 \delta -2 \Gamma  \delta  \Delta  \Delta_c+\left(\beta ^2-1\right)^2  E_{so,1}^2\delta ^2/4\notag\\
&\hspace{20pt}+\Gamma ^2 \left[\delta ^2+32 \Delta  \Delta_c-4 (\beta -1)^2  E_{so,1}\delta \right]/4\notag\\
&\hspace{40pt}- \Delta_c^2 \left[(\beta -1)^2  E_{so,1} \delta -4\Delta ^2\right]\Big\}^{1/2}.
\label{eq:interwiregeneral}
\end{align}
In the absence of  crossed Andreev pairing, the bulk gap at $k = 0$ closes at ${\bar{\Delta}}_Z^\pm$. In the regime $\sqrt{E_{so,1}\delta}/\Delta, \sqrt{\Gamma\delta}/\Delta \ll 1$ we find that
\begin{align}
&\bar{\Delta}_{Z}^- \approx \Delta - \left(\beta + 1\right)^2E_{so,1}\delta/8\Delta,
\label{eq:zeroandreev1}\\
&\bar{\Delta}_{Z}^+ \approx \sqrt{\Delta^2 + 4\Gamma^2} \label{eq:zeroandreev2} \\
&\hspace{10pt}- \left[4\Gamma\delta + \left(3\beta^2 - 2\beta+3\right) E_{so,1}\delta\right]/8\sqrt{\Delta^2 + 4\Gamma^2}.\notag
\end{align}

\begin{figure}[t!] 
\includegraphics[width=0.95\linewidth]{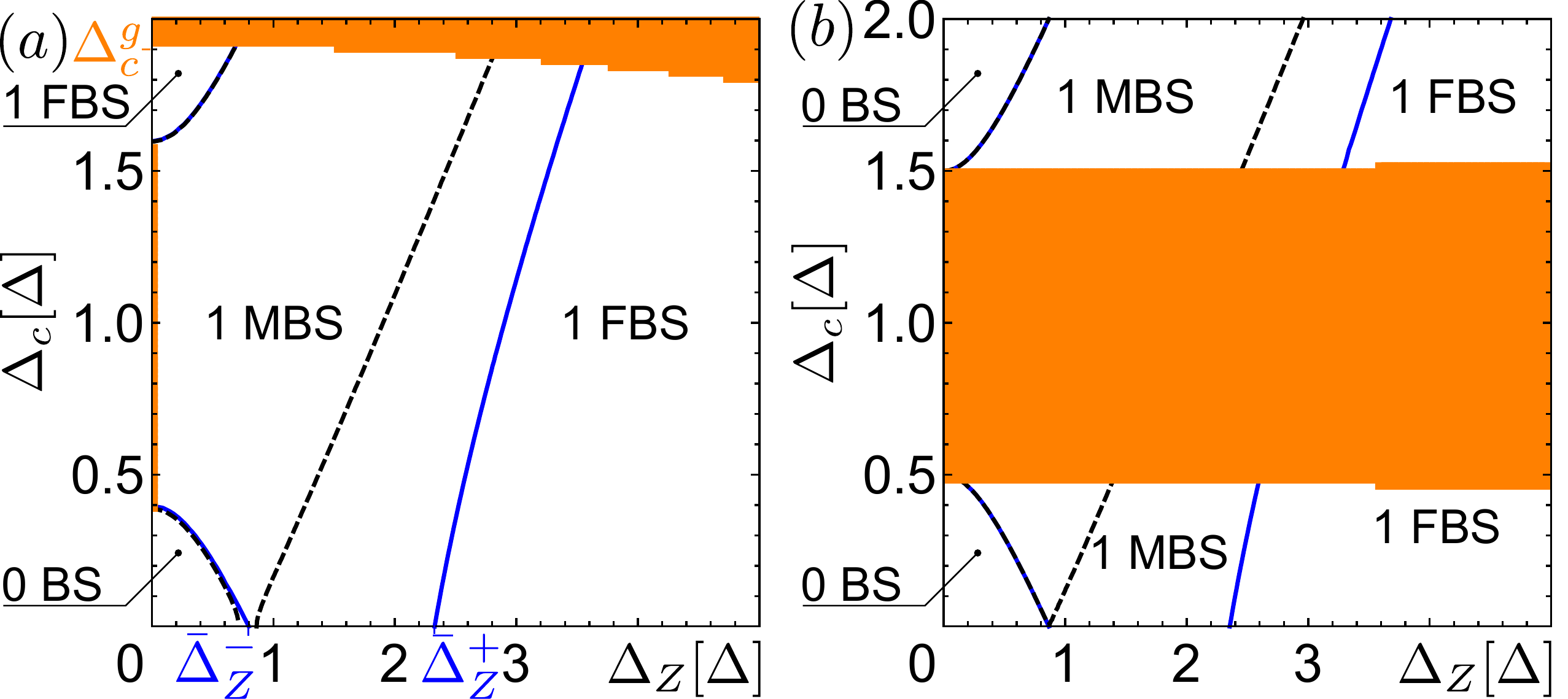}
\caption{
Phase diagram as  function of  Zeeman energy $\Delta_Z/\Delta$  and crossed Andreev pairing amplitude $\Delta_c/\Delta$. The phase boundaries  between different gapped phases for $\mu_\tau =\Gamma \equiv 0$ [$\mu_\tau =\Gamma \equiv 1.1\Delta$] are given by Eq.~\eqref{eq:phasediagram1} [Eq.~\eqref{eq:interwiregeneral}] and shown by black dashed line (blue solid line) for two cases: (a) $\beta=1.4$ and (b) $\beta=1$.
The effect of  interwire tunneling $\Gamma$ can be compensated by tuning  $ \mu_\tau$ to a sweet-spot, $ \mu_\tau = \Gamma $, for small values of $\Delta_Z$.
Note that the phase transition from the topological phase hosting one MBS at each system end to the trivial one-FBS phase is pushed to higher values of the Zeeman energy.  
For $\mu_\tau = \Gamma$ and $\Delta_c=0$, the bulk gap at $k=0$ closes at $\Delta_Z=\bar\Delta_Z^\pm$. In the finite $\Gamma$ regime [panel (a)], the gapless phase (orange area) appears at large values of  $\Delta_c^g$, while it is absent in the $\Gamma=0$ case in the same parameter range. In both cases, there is a gapless phase along $\Delta_Z = 0$ axis (orange line).
In the case  $\beta=1$ [panel (b)], a gapless phase (orange area) appears in the system both for zero and finite values of $\Gamma$ and extends over a large range of magnetic field. In contrast to the case described in  panel (a), there are no BSs in the system for $\Delta_Z < \tilde{\Delta}_{Z,-}|_{\beta=1}$ given that $\Delta_c > \Delta_c^+|_{\beta=1}$.
Other parameters are  the same as in Fig.~\ref{fig:doubleNW}. 
} 
\label{fig:diagramtunneling}
\end{figure}

We note that the prefactor in the second term in $\bar{\Delta}_{Z}^- $ is slightly modified compared to $\Delta_Z^-$ given by Eq.~\eqref{eq:zeropoints1} for $\Gamma=0$ case and since $\bar{\Delta}_{Z}^- > \Delta_Z^-$, the phase transition to a one-MBS topological phase is slightly shifted to higher values of magnetic field. By comparing Eq.~\eqref{eq:zeroandreev2} with Eq.~\eqref{eq:zeropoints2}, we note that $\bar{\Delta}_{Z}^+ \gg \Delta_Z^+$, so the phase boundary between a topological one-MBS and a trivial one-FBS phase is shifted to much higher values of the magnetic field. In general, for small values of the magnetic field the phase boundaries remain almost the same for both of the cases when $\mu_\tau=\Gamma$ is zero and finite, see Fig.~\ref{fig:diagramtunneling}(a). 
We also note that for $\mu_\tau = \Gamma$, system becomes gapless when crossed Andreev pairing $\Delta_c > \Delta_c^g$ (where $\Delta_c^g$ can be found only numerically) for  all values of Zeeman energy.  In contrast to that, as described in the Sec. \ref{sec:effectphasegradient} for zero interwire tunneling, the gapless phase is not present in the same range of parameters, see Fig.~\ref{fig:doubleNW}.

We also consider here the regime of identical NWs with equal SOI strengths, $\beta=1$. First, we discuss the phase diagram in the absence of the interwire tunneling and zero chemical potential.
As was shown in Refs. [\onlinecite{schrade2017lowfield}, \onlinecite{yanick}] for $\delta=0$, the gap in the bulk energy spectrum closes at finite momentum for all values of Zeeman energies if $\Delta_c = \Delta$.  For $\delta\neq0$, we find numerically that the energy spectrum is gapless for $\Delta_c^- \lesssim \Delta_c \lesssim \Delta_c^+$ [and all values of magnetic field  for a chosen set of parameters, see Fig. \ref{fig:diagramtunneling}(b)]. Therefore,  the region of the gapless phase is substantially larger for $\beta=1$ and it is better to work with setups that have  sufficiently different values of SOI energy in the two NWs.
Generically, the phase diagram remains almost the same as in the case $\beta>1$  discussed in Sec. \ref{sec:effectphasegradient}.  The only exception is the trivial one-FBS phase region determined by the condition $ \Delta_Z < { \tilde \Delta}_{Z,-}$ for $\Delta_c>\Delta_c^+$, which transforms into the trivial phase hosting no BSs, if $\beta=1$. 
In addition, as follows from Eqs. \eqref{eq:zeropoints1} and \eqref{eq:zeropoints2}, $\Delta_Z^- = \Delta_Z^+$ such that the bifurcation point is still present in the phase diagram and, as a result, the one-MBS phase is absent when $\Delta_c=0$. 

Next, we include the effect of the interwire tunneling on the topological phase diagram. We assume both the interwire tunneling $\Gamma$ and the chemical potentials $\mu_\tau$ to be finite and equal, $\mu_\tau = \Gamma$. Inserting $\beta = 1$ in Eq.~\eqref{eq:interwiregeneral}, we find  that the boundaries between different gapped phases are given by $\Delta_Z = \tilde{\tilde{\Delta}}_{Z,\pm}|_{\beta=1}$, where 
\begin{align}
&\tilde{\tilde{\Delta}}_{Z,-}^2|_{\beta=1} = (\Delta -\Delta_c)^2- E_{so,1} \delta  + \delta ^2 / 16,\notag\\
&\tilde{\tilde{\Delta}}_{Z,+}^2|_{\beta=1}  = (\Delta +\Delta_c)^2-  E_{so,1} \delta + (\delta - 8 \Gamma )^2 / 16.
\end{align}
Comparing this result with Eq.~\eqref{eq:phasediagram1} for the case $\beta=1$, we can see that $\tilde{\tilde{\Delta}}_{Z,-}|_{\beta=1} = 
\tilde\Delta_{Z,-}|_{\beta=1} $, while $\tilde{\tilde{\Delta}}_{Z,+}|_{\beta=1}  \gg \tilde\Delta_{Z,+}|_{\beta=1} $. Thus, the phase boundary between a one-MBS and one-FBS phase is shifted to higher values of magnetic field.
The bulk gap at $k =0$ closes at $\bar\Delta_c^\pm|_{\beta=1} $ ($\bar\Delta_Z^\pm|_{\beta=1} $) in the absence of the magnetic field (crossed Andreev pairing).  Examining the gap closing points and comparing them with the ones found for $\Gamma = 0$ [see Eqs.~\eqref{eq:zeropoints0}-\eqref{eq:zeropoints2}], we find that $\bar\Delta_c^ \pm|_{\beta=1}  = \Delta_c^ \pm|_{\beta=1} $, $\bar\Delta_Z^-|_{\beta=1}  = \Delta_Z^-|_{\beta=1} $, and $\bar\Delta_Z^+|_{\beta=1}  > \Delta_Z^+|_{\beta=1} $, where, in the regime $\sqrt{E_{so,1}\delta}/\Delta, \sqrt{\Gamma\delta}/\Delta \ll 1$,
\begin{align}
&\bar\Delta_Z^+|_{\beta=1}  \approx \sqrt{\Delta^2 + 4\Gamma^2} -\frac{(\Gamma + E_{so,1})\delta}{2\sqrt{\Delta^2 + 4\Gamma^2}}.
\end{align}
Again, we find numerically that there is a gapless phase appearing due to the gap closing at finite momentum in the bulk energy spectrum, see Fig.~\ref{fig:diagramtunneling}(b).  This gapless phase is present in both cases, namely, when interwire tunneling is zero or finite.

\subsection{Phase diagram at finite values of chemical potential}\label{sec:phasechemicalpotential}

So far, we have worked in the regime in which the chemical potentials $\mu_\tau$  in both NWs are tuned to the most optimal point ($\mu_\tau = 0$ or $\mu_\tau = \Gamma$). However, experimentally it could be quite challenging to control the position of  $\mu_\tau $ precisely, so it is important to demonstrate that the topological phase is stable against deviations of the chemical potential from the optimal value. 
Thus, in this section, we explore two parameter spaces of the topological phase diagram: $\mu_{\bar 1}$-$\Delta_Z$  and  $\mu_{\bar 1}$-$\Delta_c$.

\begin{figure}[b!] 
\includegraphics[width=0.8\linewidth]{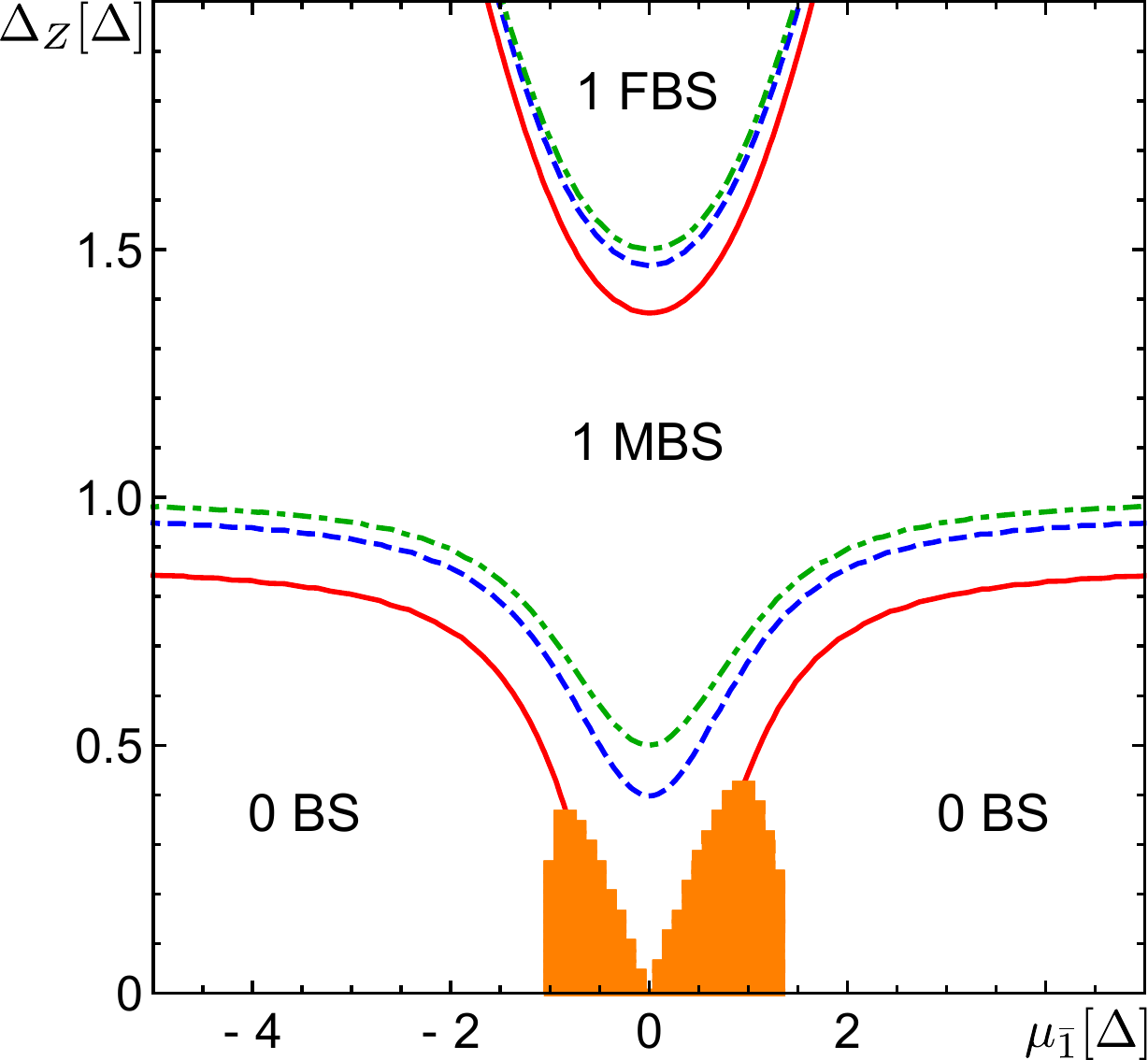}
\caption{Phase diagram of the double NW 
as  function of   chemical potential $\mu_{\bar 1}/\Delta$ and Zeeman energy $\Delta_Z/\Delta$ for different values of the phase gradient:  $\delta = 0$ (green dot-dashed line), $\delta/\Delta= 0.01$ (blue dashed line), and  $\delta/\Delta= 0.04$ (red solid line). All lines are obtained analytically from the gap closing condition at $k = 0$ given by  Eq.~\eqref{eq:phasedeltaz}.
The area corresponding to the topological phase  increases with the corresponding increase in  the supercurrent strength $\delta$, thus allowing one to apply weaker magnetic fields to achieve the one-MBS phase.  However, the topological phase at $\Delta_Z=0$ remains out of reach as the  gapless phase (orange) [shown here  for $\delta/\Delta= 0.04$]  appears at high values of $\delta$.   
The number of BSs is indicated in the plot.  
Other parameters are fixed as $\mu_1  = 0$, $\Gamma=0$, $E_{so,1}/\Delta = 6.25$, $\beta = 1.4$, and $\Delta_c/\Delta = 0.5$.} 
\label{fig:diagramchemicalpotential}
\end{figure}

\begin{figure}[t!] 
\includegraphics[width=\linewidth]{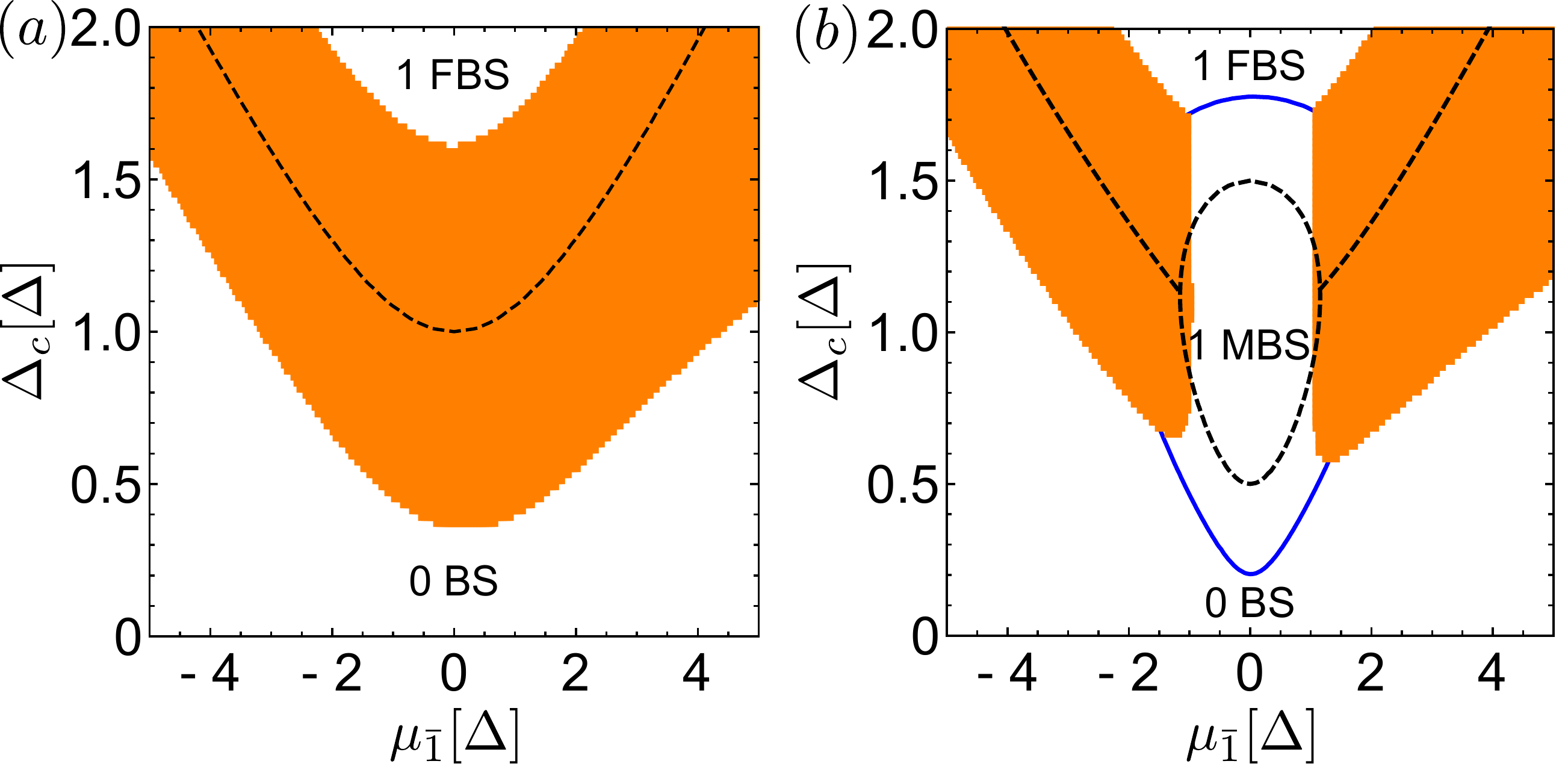}
\caption{Phase diagram of the double NW  
as  function of  chemical potential $\mu_{\bar 1}/\Delta$ and  crossed Andreev
pairing potential $\Delta_c/\Delta$. (a) For $\Delta_Z=0$, the phase boundary between the trivial and two-MBSs phase for $\delta = 0$  given by Eq.~\eqref{eq:phasezeromagnetic} is shown by the black dashed line. In the presence of supercurrent, $\delta/\Delta= 0.04$,  a gapless phase  emerges (orange region).  The one-MBS topological phase can be achieved only for finite magnetic fields.
(b) At $\Delta_Z=0.5 \Delta$, the topological phase emerges at lower values of the crossed Andreev pairing in the presence of $\delta$ [here, $\delta/\Delta= 0.04$; blue solid line is determined from the gap closing condition at $k = 0$] compared to the case $\delta = 0$ (black dashed line).
The number of BSs is indicated in the plot. 
Other parameters are chosen as $\mu_1  = 0$, $\Gamma=0$, $E_{so,1}/\Delta = 6.25$, $\beta = 1.4$, $\Delta_c/\Delta = 0.5$.} 
\label{fig:diagramzerozeeman}
\end{figure}

In the parameter space $\mu_{\bar 1}$-$\Delta_Z$,  we show that the presence of  supercurrent allows one to lower the values of the magnetic field required for entering the topological phase with one MBS also at finite chemical potential, see Fig.~\ref{fig:diagramchemicalpotential}.
For zero chemical potential in the $1$--NW, $\mu_1 = 0$, and for zero interwire tunneling $\Gamma$ (which, as shown above, can always be achieved effectively by tuning  $\mu_\tau$ to $\Gamma$), the bulk gap closes at $k = 0$, if $\Delta_Z = \Delta_{Z,\pm}$, where $\Delta_{Z,\pm}$ are real non-negative solutions of
\begin{align}
&\Delta_{Z,\pm}^2 = \Delta ^2 + \Delta_c^2 -( 1 + \beta ^2)   E_{so,1} \delta /2  +\mu_{\bar 1}^2/2 - \mu_{\bar 1}\,\delta /4 \notag\\
&\hspace{0pt} + \delta ^2/16 \pm\Big\{\Delta_c^2\left(4 \Delta ^2 +\mu_{\bar 1}^2 +2 \beta   E_{so,1} \delta \right)\notag\\
&\hspace{20pt}+ \left[\mu_{\bar 1}(\delta-2\mu_{\bar 1}) + 2(\beta^2-1) E_{so,1} \delta\right]^2/16\notag\\
 &\hspace{100pt}- (\beta^2 + 1) \Delta_c^2 E_{so,1} \delta \Big\}^{1/2}.
\label{eq:phasedeltaz}
\end{align}
For well-defined values of $\Delta_{Z,\pm}$, Eq.~\eqref{eq:phasedeltaz} implicitly gives the boundaries between different gapped phases, see Fig.~\ref{fig:diagramchemicalpotential}. 
The system is in the trivial phase without BSs, if $\Delta_Z < \Delta_{Z,-}$. If  $\Delta_{Z,-}< \Delta_Z < \Delta_{Z,+}$ the system is in the topological phase hosting one MBS at each end. If $\Delta_Z > \Delta_{Z,+}$, the system hosts one FBS.
Again, we note that, in the absence of the supercurrent, {\it i.e.} $\delta = 0$, the trivial phase with one FBS turns into the topological phase with two MBSs at each end of the system due to the presence of an additional symmetry  [\onlinecite{schrade2017lowfield}]. The minimal value of $\Delta_{Z,-}|_{\delta=0}$ to enter the topological phase is achieved, as expected, for $\mu_{\bar 1} = 0$. 
Since $\Delta_{Z,-} < \Delta_{Z,-}|_{\delta=0}$, the phase gradient lowers the critical values of the Zeeman energy for entering the topological regime with one MBS.  Moreover, $\Delta_{Z,+} < \Delta_{Z,+}|_{\delta=0}$, resulting in the trivial one-FBS phase for smaller values of magnetic field. 
We note again that if the supercurrent is too strong, the gap in the bulk  spectrum can close at finite values of the momentum, see  Fig.~\ref{fig:diagramchemicalpotential}. 
As  demonstrated previously in Figs.~\ref{fig:doubleNW} and \ref{fig:Diagramzeemanphase}, one cannot reach the topological phase in the absence of magnetic fields $\Delta_{Z}$, see Fig.~\ref{fig:diagramzerozeeman}(a). 
However, with finite supercurrents, we can shift the topological transition into a region of relatively weak magnetic field.

Next, we explore the phase diagram  in the parameter space of $\mu_{\bar 1}$-$\Delta_c$.
In the absence of supercurrent, $\delta = 0$, and of magnetic field, $\Delta_Z=0$,  the gap in the bulk spectrum closes at $k = \pm\, k_t$, 
\begin{align}
k_{t} = \dfrac{2m}{\hbar^2}\frac{\sqrt{(\beta+1)^2\alpha_1^2 + 4\mu_{\bar 1}\hbar^2/m} - (\beta+1)\alpha_1}{4 },
\label{eq:k1}
\end{align}
at the critical value of the crossed Andreev pairing $\Delta_c = \Delta_{c,t}$.
In the strong SOI regime, $E_{so,1} \gg \Delta,\mu_{\bar 1}$,  $k_t$ and $\Delta_{c, t}$ are given by
\begin{align}
&k_{t} \approx \dfrac{\mu_{\bar 1}}{(\beta+1)\alpha_1},\label{eq:k1app}\\
&\Delta_{c,t}^2 \approx \Delta ^2+\frac{\mu_{\bar 1}^2}{(\beta+1)^2} \label{eq:phasezeromagnetic}.
\end{align}

For $\Delta_c < \Delta_{c,t}$ the system is in the trivial  phase. For $\Delta_c > \Delta_{c,t}$, the system is in the topological phase hosting two MBSs protected by time-reversal symmetry. For $\mu_{\bar1} = 0$, the topological phase transition takes place for $\Delta_{c,t} = \Delta$. However, for non-zero values of $\mu_{\bar 1}$, the topological phase transition to the two-MBSs phase is shifted to higher values of $\Delta_{c,t}$ [see Fig.~\ref{fig:diagramzerozeeman}(a)]. In the presence of supercurrent, the bulk gap closes at smaller values of the crossed Andreev pairing, resulting in a lower threshold to leave the trivial phase compared to the case without supercurrent.
However, the system, instead of entering the topological phase, remains gapless. As a result, the system develops a bulk gap only at high values of $\Delta_c > \Delta_{c,t}$, for which it enters a trivial phase hosting one FBS at each system end,
 see Fig.~\ref{fig:diagramzerozeeman}(a).

To achieve the topological phase hosting one MBS, one needs to apply  a magnetic field, see Fig. \ref{fig:diagramzerozeeman}(b). In the presence of supercurrent, similar to the previous sections, the threshold of the magnetic field and the crossed Andreev pairing to enter  the topological phase can be lowered, see also Figs.~\ref{fig:doubleNW}, \ref{fig:Diagramzeemanphase}, and \ref{fig:diagramchemicalpotential}.
However, generally, the system is gapless
for a large range of  parameters.
For $\delta = 0$, the gap in the bulk  spectrum closes at zero momentum provided that $\Delta_c = \tilde{\Delta}_{c,t}$,  $\tilde{\Delta}_{c,t}^2 = \Delta ^2+\Delta_Z^2 \pm \sqrt{\mu_{\bar 1}^2 \left(\Delta_Z^2-\Delta ^2\right)+4 \Delta ^2 \Delta_Z^2}$, determining the boundaries between different gapped phases.
Again, the gap closing at finite values of the momentum  can be found only numerically.
In the presence of magnetic fields, the effect of supercurrent is less damaging such that the gapless phase is not present at small values of $\mu_{\bar 1}$.  Importantly, the threshold for the amplitude of the crossed Andreev pairing required for entering the one-MBS topological phase is decreased compared to the $\delta=0$ case,
see Fig.~\ref{fig:diagramzerozeeman}(b). Furthermore, the upper threshold in $\Delta_c$ to leave the topological one-MBS phase and to enter into the trivial one-FBS phase is increased, which results in an enlarged regime of topological phase. 
Thus, the presence of the supercurrent  is again shown to be beneficial for realizing the one-MBS phase.

\section{Conclusions} \label{sec:conclusion}
We studied a double-NW setup consisting of two parallel NWs with Rashba SOI proximitized with a supercurrent-carrying bulk  $s$-wave superconductor and subjected to a magnetic field.  In the presence of the supercurrent, the direct $\Delta$ and  crossed Andreev $\Delta_c$  pairing amplitudes acquire a phase gradient, which serves as an additional parameter to tune the system into the topological phase. Importantly, the system enters  the topological phase, characterized by the presence of one MBS at each system end, at lower values of the magnetic field compared to standard single-NW setup or double-NW setup without phase gradient. Thus, such supercurrents in combination  with crossed Andreev pairing 
allows one to work at smaller magnetic fields, making the double-NW setup more attractive than a single-NW one.

\emph{Acknowledgments.} This work was supported by the Swiss National Science Foundation and  NCCR QSIT. This project received funding from the European Union's Horizon 2020 research and innovation program (ERC Starting Grant, grant agreement No 757725).


\begin{thebibliography}{99}
\bibitem{noSOI1} We note that, in our setup, SOI plays an important role. The double-NW setup without SOI was considered in Ref.~[\onlinecite{noSOI}].
\end{thebibliography}
\end{document}